\documentclass{article}

\usepackage{microtype}
\usepackage{graphicx}
\usepackage{subfigure}
\usepackage{booktabs} % for professional tables

\usepackage{hyperref}

\usepackage[accepted]{icml2025}

\usepackage{amsmath}
\usepackage{amssymb}
\usepackage{mathtools}
\usepackage{amsthm}

\usepackage{listings}
\usepackage{xcolor}

\lstdefinelanguage{json}{
  basicstyle=\ttfamily\small,
  numbers=left,
  numberstyle=\tiny\color{gray},
  stepnumber=1,
  numbersep=5pt,
  showstringspaces=false,
  breaklines=true,
  frame=single,
  backgroundcolor=\color{gray!5},
  literate=
   *{0}{{{\color{blue}0}}}{1}
    {1}{{{\color{blue}1}}}{1}
    {2}{{{\color{blue}2}}}{1}
    {3}{{{\color{blue}3}}}{1}
    {4}{{{\color{blue}4}}}{1}
    {5}{{{\color{blue}5}}}{1}
    {6}{{{\color{blue}6}}}{1}
    {7}{{{\color{blue}7}}}{1}
    {8}{{{\color{blue}8}}}{1}
    {9}{{{\color{blue}9}}}{1}
    {:}{{{\color{black}:}}}{1}
    {,}{{{\color{black},}}}{1}
    {"}{{{\color{red}"}}}{1}
}

\usepackage{tcolorbox}
\tcbuselibrary{listingsutf8}

\usepackage[capitalize,noabbrev]{cleveref}

\theoremstyle{plain}

\theoremstyle{definition}

\theoremstyle{remark}

\usepackage[textsize=tiny]{todonotes}

\icmltitlerunning{Submission and Formatting Instructions for ICML 2025}

\begin{document}

\twocolumn[
\icmltitle{AGACCI : Affiliated Grading Agents for Criteria-Centric Interface\\ in  Educational Coding Contexts
}



\icmlsetsymbol{equal}{*}

\begin{icmlauthorlist}
\icmlauthor{Kwangsuk Park}{equal,aalab,aiffel}
\icmlauthor{Jiwoong Yang}{equal,aalab,inha}
\end{icmlauthorlist}

\icmlaffiliation{aalab}{AA LAB, MODULABS}
\icmlaffiliation{aiffel}{Aiffel, MODULABS}
\icmlaffiliation{inha}{Department of Statistics, Inha university}

%\icmlaffiliation{yyy}{Department of XXX, University of YYY, Location, Country}
%\icmlaffiliation{sch}{School of ZZZ, Institute of WWW, Location, Country}

\icmlcorrespondingauthor{Kwangsuk Park}{KSPARK@MODULABS.CO.KR}
\icmlcorrespondingauthor{Jiwoong Yang}{DIDWLDND960923@GMAIL.COM}

\icmlkeywords{Multi Agents, ICML}

\vskip 0.3in
]



\printAffiliationsAndNotice{\icmlEqualContribution} % otherwise use the standard text.

\begin{abstract}
Recent advances in AI-assisted education have encouraged the integration of vision-language models (VLMs) into academic assessment, particularly for tasks that require both quantitative and qualitative evaluation. However, existing VLM-based approaches struggle with complex educational artifacts, such as programming tasks with executable components and measurable outputs, that require structured reasoning and alignment with clearly defined evaluation criteria. We introduce AGACCI, a multi-agent system that distributes specialized evaluation roles across collaborative agents to improve accuracy, interpretability, and consistency in code-oriented assessment. To evaluate the framework, we collected 360 graduate-level code-based assignments from 60 participants, each annotated by domain experts with binary rubric scores and qualitative feedback. Experimental results demonstrate that AGACCI outperforms a single GPT-based baseline in terms of rubric and feedback accuracy, relevance, consistency, and coherence, while preserving the instructional intent and evaluative depth of expert assessments. Although performance varies across task types, AGACCI highlights the potential of multi-agent systems for scalable and context-aware educational evaluation.
\end{abstract}

\section{Introduction}
\label{submission}

The integration of generative AI into educational settings has emerged as a promising avenue to improve instructional quality and alleviate teacher workload through scalable and personalized support. In particular, generative models hold strong potential for enhancing student assessment by delivering real-time, individualized feedback tailored to the semantic content and structure of student work. However, despite this promise, current large language model (LLM)-based assessment systems exhibit several critical limitations. Generated feedback often fails to reflect students’ performance or misconceptions, instead producing superficial praise or vague advice. In many cases, a single LLM can generate overly positive evaluations for incorrect responses or include hallucinated reasoning lacking evidential basis \cite{jansen2024empirische}. In addition, existing automated evaluation methods frequently neglect the fine-grained criteria defined in instructional rubrics, focusing instead on surface-level correctness or grammar \cite{phung2023generative}. This misalignment can result in feedback that diverges from the instructor's intent or omits key evaluative dimensions. Lastly, LLM-based judgments have been observed to vary inconsistently among similar or identical responses. Although ensemble-based strategies have been proposed to address such instability \cite{pathak2025rubric}, the structural limitations of single-model systems continue to pose challenges for reliability in automated feedback.

To address these limitations, a variety of approaches have been proposed. G-Eval \cite{liu2023g} employs large language models as evaluators to assess the quality of generated outputs without relying on reference answers, achieving high alignment with human judgment. More recent studies have extended this idea through generate–evaluate–regenerate pipelines, improving the reliability and validity of automated feedback \cite{guo2024using}; \cite{seo2025large}. And recent method \cite{zhuge2024agent} shows that dedicating a separate agent solely to evaluation can further boost alignment with human graders. Nonetheless, such systems remain structurally constrained: error detection is often imperfect, and identified issues are not always fully reflected in the final feedback. Furthermore, although multi-criteria frameworks have been introduced to improve feedback quality, challenges persist in ensuring consistency across evaluation dimensions and in aligning feedback with fine-grained rubric components. These limitations suggest that more systematic role allocation and structured evaluation pipelines may offer a viable path forward.

\begin{figure}[ht]
\begin{center}
  \centerline{\includegraphics[width=0.85\columnwidth]{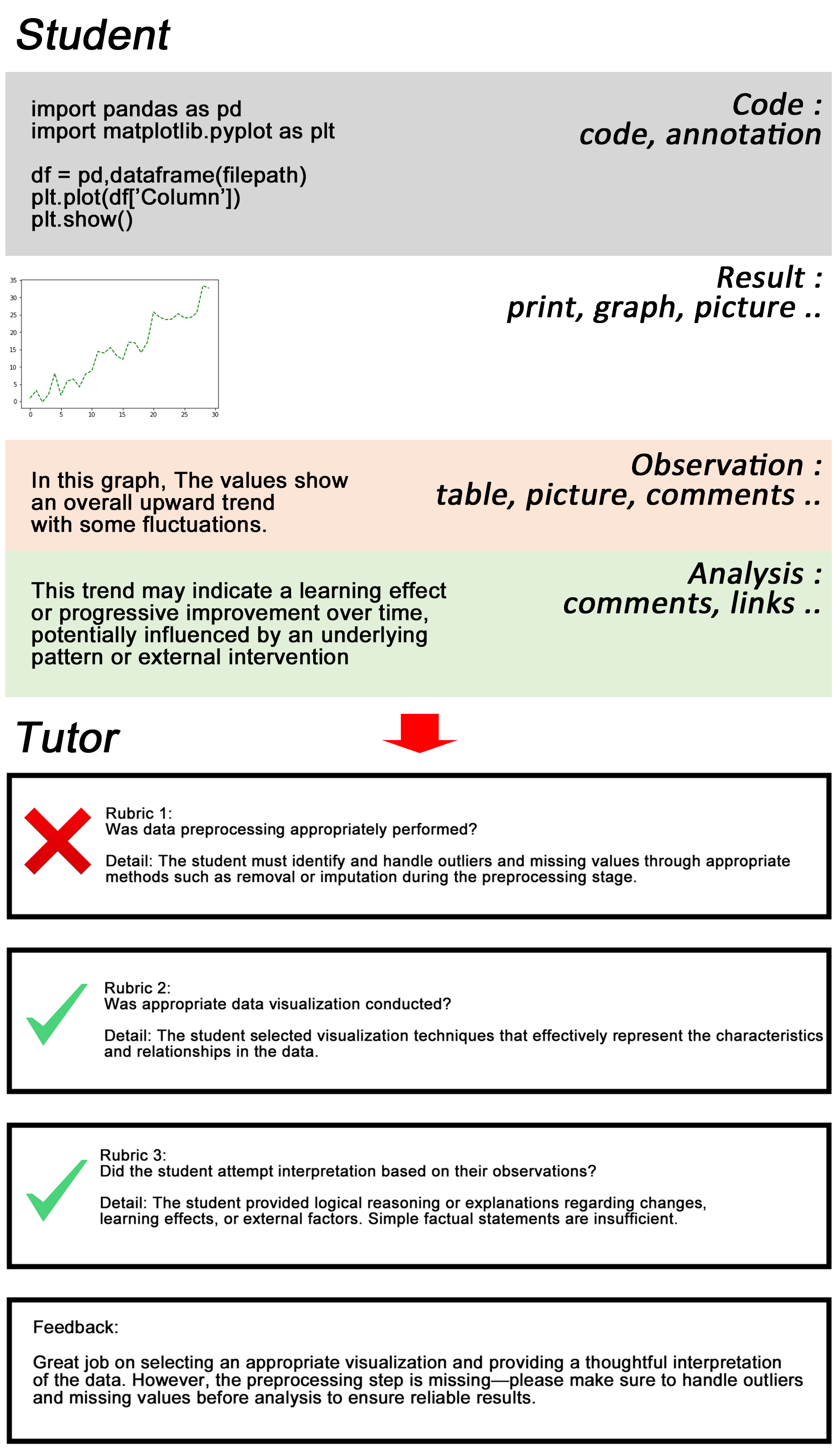}}
  \caption{An example of the multi-faceted student submissions AGACCI is designed to evaluate, involving code execution, data visualization, and interpretive reasoning. The figure illustrates how rubric-based assessment decomposes these components into distinct evaluation criteria, supporting targeted and interpretable feedback.}
\end{center}
\end{figure}

A recent modular framework \cite{lee2024vista} demonstrated that separating task-specific LLM components can improve educational content generation. Building on this insight, to provide accurate, consistent, and interpretable automated educational assessment in a principled and scalable manner, we propose AGACCI (Affiliated Grading Agents for Criteria-Centric Interface), a multi-agent evaluation framework. AGACCI is designed to assess complex student submissions—comprising code, graphs, markdowns and other multimodal artifacts —that written in Jupyter Notebooks, widely adopted in programming education for their ability to combine executable code, visual output, and explanatory text within a single interactive environment \cite{kluyver2016jupyter}, and align with rubric-based evaluation goals specified by domain experts. The framework decomposes the evaluation process into modular roles, including code execution and result analysis, visualization assessment, and interpretive reasoning, thereby systematically addressing issues observed in prior systems such as criterion misalignment, insufficient error reflection, and lack of analytical depth. Each agent is specialized in a distinct evaluation function, while a meta-evaluator ensures the internal coherence and evidential basis of the system’s outputs. The final judgments and feedback are synthesized through an integrative decision process. This architecture enables AGACCI to deliver multifaceted analysis and expert-level evaluations of complex, code-centric student work.

We evaluate AGACCI on a dataset comprising 360 code-based submissions, collected from 60 participants who completed six identical programming tasks. Each submission was annotated by domain experts using three binary rubric scores and qualitative feedback. Compared to a single GPT-based evaluator, AGACCI consistently outperforms the baseline in rubric alignment, feedback consistency, and interpretive depth. In addition, it generates feedback that is more rubric-grounded, specific, and pedagogically meaningful that enabling clearer instructional guidance and richer formative support for learners.

Our contributions are three-fold:
\begin{itemize}
\item We present AGACCI, a modular multi-agent framework that operationalizes rubric-based evaluation through distributed roles, enabling structured assessment of complex student work involving code, visualizations, and reasoning.
\item We show that AGACCI delivers clearer, more criterion-aligned feedback than single-model systems by structurally decomposing the evaluation process into specialized agents. This design not only improves evaluative accuracy and alignment with instructional rubrics, but also enhances the pedagogical value of feedback by making it more specific, actionable, and reflective of student performance.
\item We highlight a viable path toward AI-integrated classroom assessment by demonstrating AGACCI’s potential to augment teacher capacity and improve the scalability of formative feedback, as validated through expert-annotated submissions collected from a real-world educational setting.
\end{itemize}

\section{AGACCI}

To provide accurate and rubric-aligned feedback on student code, we developed a multi-agent system where each agent is assigned a specialized evaluation role. These agents work in coordination, with each one focusing on a distinct aspect of the assessment process—such as execution, interpretation, or alignment. This division of responsibility enables the system to efficiently analyze submissions across multiple dimensions, resulting in more reliable, informative, and pedagogically meaningful feedback.
As illustrated in Figure 2, the system is composed of a pipeline of modular agents, each responsible for a distinct stage in the evaluation process. The workflow begins with rubric interpretation and submission analysis, which decompose the task into structured evaluation targets. The system then branches into three parallel evaluation streams: execution-based assessment (via the Execution and Result Evaluators), visual output assessment (via the Visualization Evaluator), and reasoning assessment (via the Interpretation Evaluator). These specialized agents operate concurrently to analyze different dimensions of the submission. A Meta Evaluator ensures cross-stream consistency, and a Final Judge integrates all outputs into a rubric-aligned decision, which is then formatted by the Summarizer for feedback delivery. This architecture supports both parallelism and hierarchical control, enabling robust, interpretable, and scalable assessment of complex student work.

\begin{figure}[ht]
\begin{center}
  \centerline{\includegraphics[width=0.85\columnwidth]{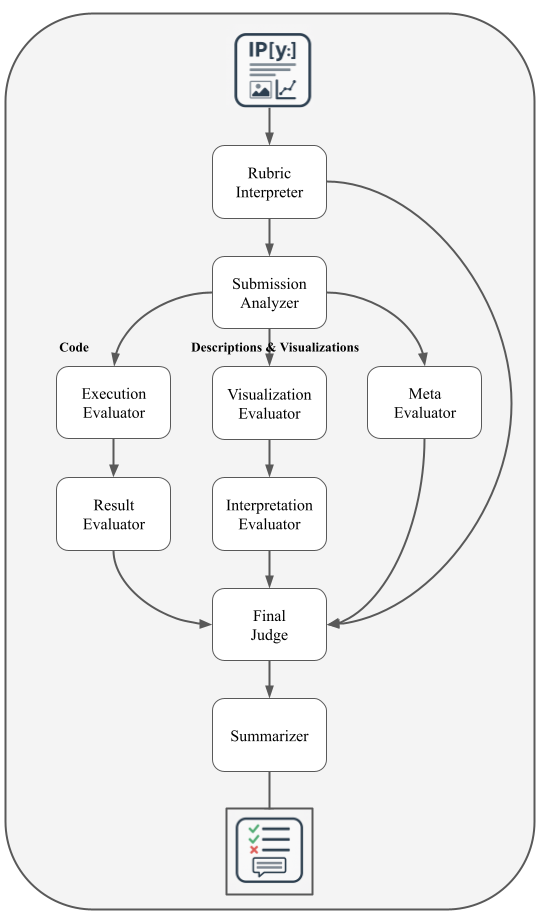}}
  \caption{Figure 2. Architecture of AGACCI. The system parses rubrics and student submissions, then branches into three evaluation paths: execution and result analysis, visualization assessment, and interpretation analysis. A meta-evaluator checks consistency, and the Final Judge aggregates all outputs. The Summarizer generates structured feedback.}
\end{center}
\end{figure}

\paragraph{\textit{Rubric Interpreter}} The Rubric Interpreter agent parses high-level rubric descriptions and restructures them into operational criteria. Rather than treating rubrics as static checklists, this agent transforms them into actionable evaluation goals by identifying implicit dependencies, ordering constraints, and minimum performance expectations. Its output enables downstream evaluators to operate with clarity and alignment to educational intent.

\paragraph{\textit{Submission Analyzer}} The Submission Analyzer agent reviews the student's submission holistically, identifying its main objectives, logical structure, and alignment with rubric criteria. It detects the sequence and purpose of code blocks, commentary, and outputs, acting as a bridge between human-readable goals and machine-level analysis. The Submission Analyzer ensures that downstream evaluations occur within the appropriate pedagogical context.

\paragraph{\textit{Execution Evaluator}} The Execution Evaluator focuses on the functional validity of the student's code. It checks whether the code runs without errors, whether core computational steps are present, and whether expected outputs (e.g., plots or printed metrics) are generated. This agent ensures the reliability of technical performance before qualitative assessment begins.

\paragraph{\textit{Result Evaluator}} The Result Evaluator agent determines whether the executed code meets the quantitative performance criteria defined in the rubric—such as accuracy thresholds, leaderboard scores, or other output metrics. It parses printed outputs, logs, or numerical results to assess success or failure. If execution fails or produces no measurable result, this agent awaits instruction and abstains from judgment. Its evaluation is binary (pass/fail), with justification grounded in observable, rubric-aligned metrics.

\paragraph{\textit{Visualization Evaluator}} The Visualization Evaluator examines the clarity and appropriateness of visual outputs, such as charts or graphs. It considers whether the selected visualization methods match the data's nature and whether the visual components (e.g., axes, labels, legends) support interpretability. This agent judges whether the output can effectively communicate insights to learners and reviewers alike.

\paragraph{\textit{Interpretation Evaluator}} The Interpretation Evaluator agent assesses the student’s ability to reason beyond observation. It looks for causal or inferential explanations that draw meaning from patterns, anomalies, or trends in the data or visualizations. It flags overly descriptive or insufficiently justified comments, rewarding interpretations that demonstrate depth and logical rigor.

\paragraph{\textit{Meta Evaluator}} Operating as an internal consistency checker, the Meta Evaluator cross-validates agent outputs, flagging contradictions or unsupported assessments. It performs alignment checks between observed evidence and declared rubric satisfaction, and may suggest overrides or confidence adjustments before final grading occurs.

\paragraph{\textit{Final Judge}} The Final Judge agent aggregates all evaluations into a conclusive decision. It resolves ambiguities across agent outputs and determines binary rubric satisfaction scores (pass/fail). This agent also generates human-readable feedback that highlights both strengths and shortcomings, balancing fairness with pedagogical value.

\paragraph{\textit{Summarizer}} The Summarizer consolidates the system's verdict into a compact, learner-facing summary. It extracts essential findings, recommendations, and rubric scores into a structured format (e.g., JSON), suitable for logging, display, or human review. The Summarizer ensures that feedback is not only accurate, but also concise and comprehensible.

To support agent-level coordination and flexible interaction patterns, we built our system on top of AutoGen \cite{wu2023autogen}, an open-source framework for constructing multi-agent applications powered by large language models. This infrastructure enables modular agent orchestration, but its effectiveness in practice largely depends on the underlying language model’s ability to support efficient, scalable interaction.

Considering the constraints of real-world educational environments—such as limited computational resources, budget constraints, and the need for responsive feedback—we selected GPT-4o mini as the backbone model for AGACCI. While smaller in scale than full-sized foundation models, GPT-4o mini provides sufficient reasoning performance for pedagogically relevant tasks, including rubric-aligned evaluation, code analysis, and interpretive feedback generation. Its lightweight architecture enables faster, more consistent inference, which is critical for maintaining interactivity and scalability in classroom settings. Furthermore, its efficiency allows seamless integration within AGACCI’s multi-agent structure, supporting parallel agent deployment without excessive computational overhead. These characteristics make GPT-4o mini a practical and sustainable choice for classroom-aligned automated assessment systems.

\section{Evaluation}

\subsection{Dataset}
To evaluate the performance of AGACCI, we conducted experiments using student submissions collected from an actual university-level course setting. A total of 60 participants completed six graduate-level programming assignments, each focused on a distinct domain within the AI curriculum: machine learning, computer vision, and natural language processing. Each assignment was accompanied by a domain-specific rubric developed by instructional staff, consisting of three clearly defined evaluation criteria. These rubric items reflected both technical correctness and interpretive reasoning, providing a fine-grained basis for automated assessment. By applying the same rubric structure consistently across all 360 collected submissions, we ensured a standardized evaluation framework that supports reliable comparison between systems.

\subsection{Evaluation Strategy}
To assess the quality and reliability of feedback generated by AGACCI, we designed a dual evaluation strategy comprising both qualitative and quantitative measures. Given the inherent stochasticity of VLM outputs—where identical prompts may yield varying results—we conducted six independent evaluation rounds for both the baseline single-model approach (SLI) and the AGACCI framework. To ensure a fair and capacity-matched comparison, we used GPT-4o mini as the underlying model for both systems.

In each round, the model received the same input, consisting of the student’s code and the corresponding rubric, and produced a feedback comment evaluating rubric achievement. The generated comments were compared against expert-written reference feedback created by AI practitioners familiar with the rubric. These references served as ground truth, offering explicit diagnoses of student performance and rubric satisfaction. This structure enabled us to evaluate both the semantic quality of the feedback and its alignment with pedagogical goals.

For qualitative evaluation, we employed an LLM-assisted rubric-aligned assessment protocol inspired by G-Eval. Feedback quality was assessed across four dimensions:

\paragraph{\textit{Feedback Accuracy}} This criterion evaluates how accurately the feedback reflects the code’s purpose, structure, and logic. Comments with vague praise or superficial descriptions receive lower scores, while those that clearly explain the code’s intent, flow, and operations are rewarded. The prompt encourages penalizing generic or underspecified responses, even if broadly accurate, to emphasize technical specificity.
    
\paragraph{\textit{Relevance to Rubrics}} This dimension evaluates how well the feedback aligns with the three rubric items defined for each assignment. Comments that fail to address rubric criteria or rely on generic language receive lower scores. Rather than offering general code assessment, this criterion rewards feedback that explicitly discusses rubric satisfaction and provides justification. The prompt is designed to penalize omissions or misinterpretations of rubric elements, quantifying the model’s ability to internalize rubric-based evaluation.
    
\paragraph{\textit{Consistency}} This criterion assesses whether the feedback maintains a coherent evaluative stance throughout the comment, avoiding internal contradictions or shifts in reasoning. For instance, praising performance initially and later criticizing it without justification is considered inconsistent. This reflects the need for logical flow and stable judgment, which are crucial in instructional contexts.
    
\paragraph{\textit{Coherence}} Coherence refers to the overall structural integrity and conceptual organization of the feedback. Beyond sentence-level fluency, this criterion values logical progression and integration of ideas. Feedback that consists of well-structured reasoning rather than a disjointed list of points is rated higher, reflecting the model’s ability to produce pedagogically meaningful and cohesive commentary.

Each feedback comment was evaluated on a 5-point scale. To reduce scoring bias and mitigate the variance inherent in LLM outputs, we repeated the G-Eval prompt 20 times per comment, using the average score as the final evaluation. This approach improves the stability and reproducibility of automated assessment, addressing the limitations of single-response evaluation.

For quantitative evaluation, we extracted binary rubric decisions (0 or 1) from the generated comments and compared them to the expert-assigned rubric scores. Since each assignment comprises three rubric items, this task was framed as a multi-label binary classification problem, with average rubric accuracy across submissions serving as the primary metric.

Since all experimental artifacts were written in Korean, we employed GPT-4o as the scoring model for evaluating feedback quality across all criteria. GPT-4o ranks among the top-performing models for Korean language tasks on the Chatbot Arena Leaderboard \cite{chiang2024chatbot}, reflecting its strong capabilities in Korean understanding and generation. This made it a reliable and linguistically appropriate reference model for our evaluation framework.

\section{Result}

\begin{figure}[ht]
\begin{center}
  \centerline{\includegraphics[width=\columnwidth]{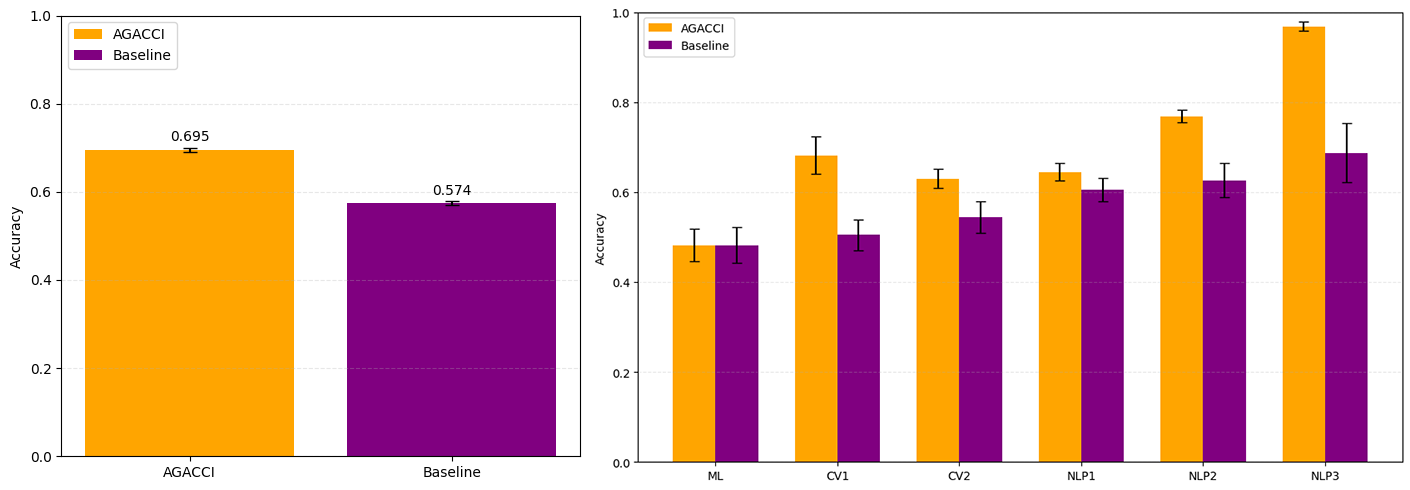}}
  \caption{Left: Overall rubric accuracy of AGACCI and the single-model baseline. AGACCI shows a substantial improvement over the baseline. Right: Accuracy per assignment domain. AGACCI consistently outperforms the baseline across domains, with particularly large gains in NLP tasks.}
\end{center}
\end{figure}

\subsection{Quantitive result}

AGACCI consistently outperformed the single-model baseline in rubric-level classification accuracy. As shown in Figure 3 (left) , AGACCI achieved 12 percentage higher average accuracy compared to the baseline. TThis performance gain highlights the advantage of AGACCI’s structured, role-based evaluation approach in aligning with expert-assigned rubric judgments, particularly by allowing specialized agents to focus on distinct evaluation dimensions and minimizing conflicting or underspecified interpretations in the final feedback.

Figure 3 (right) presents rubric-level accuracy scores across six assignment domains spanning, regression task referred as ML, computer vision tasks (face detection as CV1, segmentation as CV2), and natural language processing tasks (text classification as NLP1, summarization as NLP2, and chatbot interaction as NLP3). AGACCI consistently outperformed the single-model baseline across domains, with particularly strong gains observed in NLP tasks. These tasks involve substantial interpretive reasoning and coherent language generation—capabilities that benefit from AGACCI’s structured multi-agent evaluation process. The framework’s modular design enables each agent to focus on a distinct evaluative function, resulting in more rubric-aligned judgments and improved accuracy across both technical and interpretive dimensions of student work.

While AGACCI outperforms the baseline in most domains, its performance in the ML assignment is notably comparable to the baseline, particularly for rubric items 2 and 3. We attribute this to the nature of these rubrics in Table 1, which require evidence of external actions—such as successful Kaggle submission and leaderboard threshold verification—that cannot be directly inferred from the code itself. Since AGACCI relies solely on the code and rubric context to evaluate submissions, it may fail to detect these external outcomes unless explicitly documented by the student. This highlights a structural limitation of code-centric evaluation systems in handling criteria dependent on post-submission actions or external validation.

\subsubsection{Fine-Grained Analysis by Rubric Criteria}

\begin{table}[t]
\caption{Rubric items of ML task}
\label{sample-table}
\vskip 0.15in
\begin{center}
\begin{small}
\begin{sc}
\resizebox{\columnwidth}{!}{
\begin{tabular}{lcccr}
\toprule
Rubric Item \\
\midrule
Preprocessing, training, and visualization \\
Kaggle submission status \\
Leaderboard accuracy threshold met \\
\bottomrule
\toprule
AGACCI & SLI \\
\midrule
0.7337$\pm$0.0978 & 0.1739$\pm$0.0178 \\
0.4728$\pm$0.0109 & 0.5870$\pm$0.0589 \\
0.2391$\pm$0.0000 & 0.6848$\pm$0.0416 \\
\bottomrule
\end{tabular}
}
\end{sc}
\end{small}
\end{center}
\vskip -0.1in
\end{table}

\begin{table}[t]
\caption{Selected rubric items with large performance gaps}
\label{rubrics}
\vskip 0.15in
\begin{center}
\begin{small}
\begin{sc}
\resizebox{\columnwidth}{!}{
\begin{tabular}{lcccr}
\toprule
Task & Rubric Item \\
\midrule
ML & Preprocessing, training, and visualization \\
CV1 & Natural sticker alignment on face \\
CV2 & Solution for portrait mode errors \\
CV2 & Clear identification of portrait errors \\
NLP1 & Improved accuracy using Word2Vec \\
NLP2 & Extractive vs. abstractive comparison \\
NLP3 & Stable Transformer convergence \\
NLP3 & Korean response generation model \\
\bottomrule
\toprule
AGACCI & SLI \\
\midrule
0.7337$\pm$0.0978 & 0.1739$\pm$0.0178 \\
0.7455$\pm$0.0268 & 0.1786$\pm$0.0484 \\
0.6798$\pm$0.0439 & 0.3860$\pm$0.0516 \\
0.6535$\pm$0.0088 & 0.4342$\pm$0.0461 \\
0.6509$\pm$0.0189 & 0.4057$\pm$0.0109 \\
0.8673$\pm$0.0204 & 0.4541$\pm$0.0510 \\
0.9694$\pm$0.0204 & 0.5765$\pm$0.0918 \\
0.9592$\pm$0.0000 & 0.5102$\pm$0.0957 \\
\bottomrule
\end{tabular}
}
\end{sc}
\end{small}
\end{center}
\vskip -0.1in
\end{table}

As shown in Table 2, AGACCI demonstrated clear performance advantages in rubric items that required not just binary judgments, but layered structural understanding and multi-step reasoning. Notably, it outperformed the single-model baseline in tasks involving visual coherence and explanatory adequacy (CV1), error diagnosis and solution proposal (CV2), comparative summarization strategies (abstractive vs. extractive, NLP2), and stable implementation of deep learning models (NLP3). In these high-complexity evaluation targets, AGACCI maintained an average rubric-level accuracy above 0.73, significantly exceeding that of the baseline.

These results suggest that AGACCI is particularly well-suited for assessments demanding nuanced interpretation, conditional reasoning, and inter-agent coordination. Its multi-agent architecture enables it to reconstruct structural logic and provide coherent explanations when rubric criteria involve visual layout, model behavior, or learning process articulation. Accordingly, AGACCI offers a more appropriate evaluation paradigm for future educational settings where interpretive depth and explanatory rigor are central to assessment.

\subsection{Qualitative result}

\begin{figure}[ht]
\begin{center}
  \centerline{\includegraphics[width=0.6\columnwidth]{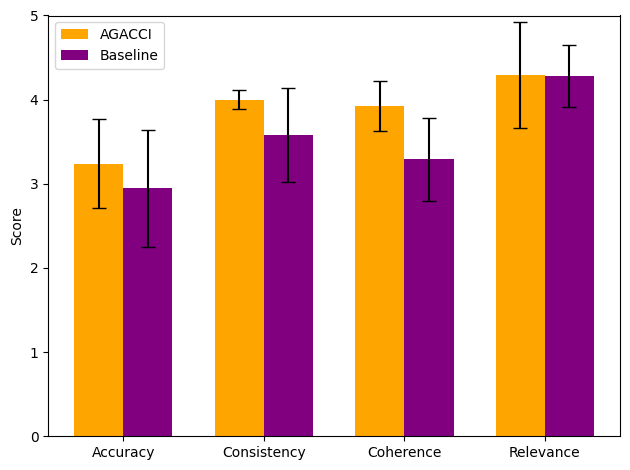}}
  \caption{Scores comparing AGACCI and the single-model baseline across four qualitative criteria: Feedback Accuracy, Consistency, Coherence, and Relevance. AGACCI outperforms the baseline in all dimensions except Relevance, where both systems perform comparably. }
\end{center}
\end{figure}

As shown in Figure 4 and 5, AGACCI consistently surpassed the single-model baseline across multiple qualitative dimensions, notably in consistency and coherence. In terms of consistency, AGACCI achieved both a higher mean score and a markedly lower variance. This indicates that its feedback maintained a more stable evaluative stance across responses, avoiding contradictions or shifts in judgment that were more frequently observed in the baseline outputs. This stability is likely attributable to the Meta Evaluator, a coordination agent designed to reconcile conflicting assessments across specialized modules. By explicitly checking alignment among agents’ outputs before finalizing feedback, AGACCI is able to preserve logical continuity and evaluative integrity in its comments.

In addition to consistency, AGACCI outperformed the baseline in coherence, producing feedback that exhibited stronger internal structure and integrated reasoning. This improvement can be traced to the system’s modular architecture, which assigns agents to specialized roles (e.g., execution, interpretation, visualization), allowing each to construct a precise sub-evaluation before aggregation. The result is feedback that not only touches on relevant evaluation criteria, but also weaves them into a coherent narrative. In pedagogical terms, this coherence enhances the learner’s ability to understand the rationale behind the evaluation, improving both clarity and receptiveness to feedback.

While rubric relevance scores were high for both systems, AGACCI demonstrated slightly more stable performance with reduced variance. This suggests that its rubric interpretation and alignment mechanisms, especially the Rubric Interpreter agent, consistently map high-level rubric language into actionable, structured targets across different prompts. As a result, AGACCI is less prone to omitting rubric-aligned feedback or drifting into generic comments—a common failure mode of single-model systems when rubrics are implicit or abstract.

Taken together, these findings highlight the practical value of distributing evaluative responsibilities across dedicated agents. AGACCI’s architecture not only enhances technical feedback accuracy but also enables high-level reasoning, consistency, and communicative quality in generated feedback. Such characteristics are critical for reliable automated assessment in educational contexts, where transparency and instructional value are just as important as correctness.

\begin{figure}[ht]
\begin{center}
  \centerline{\includegraphics[width=\columnwidth]{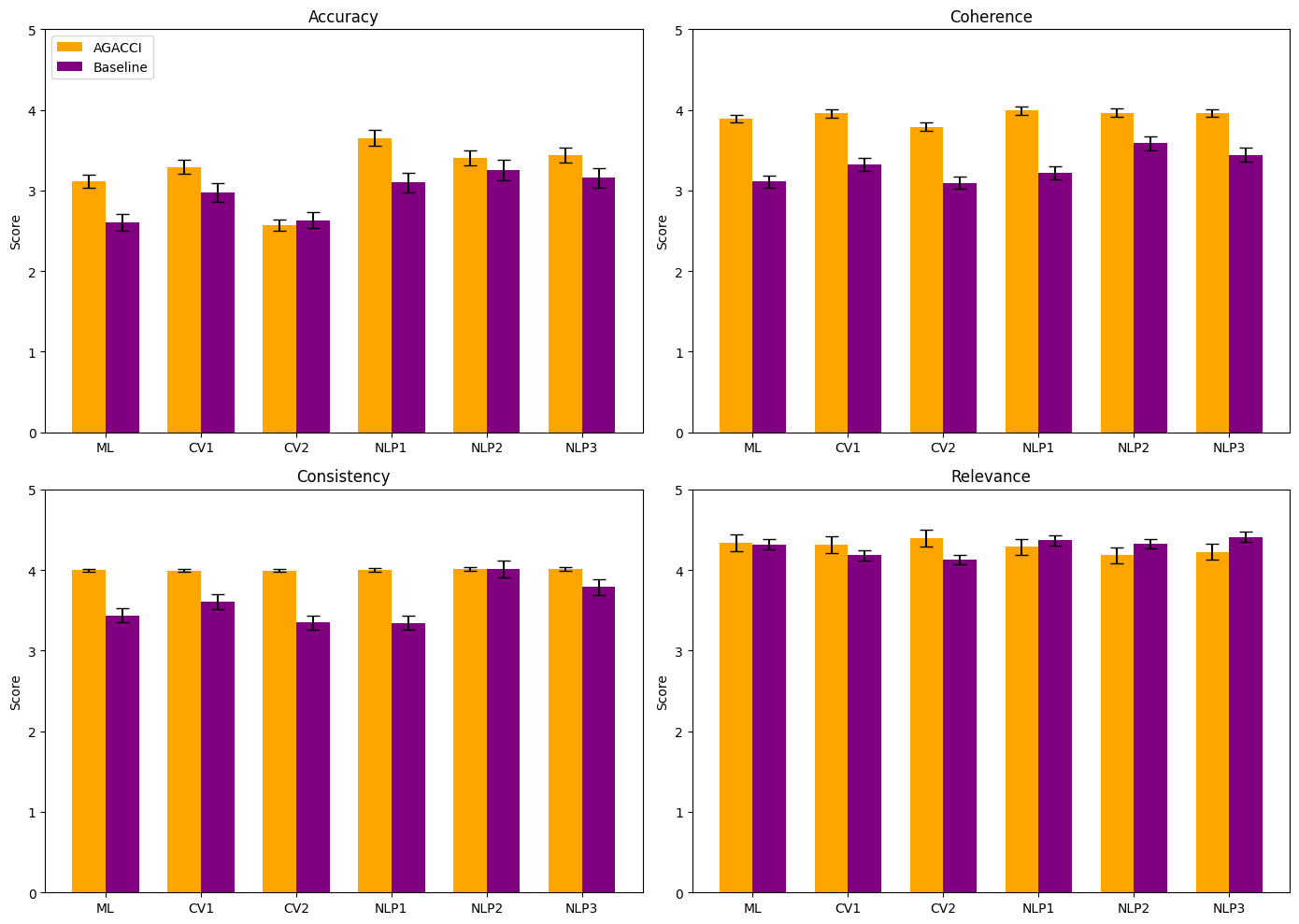}}
  \caption{Comparison of AGACCI and the single-model baseline across four qualitative evaluation criteria: Feedback Accuracy, Coherence, Consistency, and Relevance, measured over six assignment categories. AGACCI consistently achieves higher scores in Accuracy, Coherence, and Consistency across domains, with particularly strong performance in NLP tasks. Both systems perform comparably in Relevance.}
\end{center}
\end{figure}

\begin{table}[ht]
\caption{Example feedback from AGACCI, SLI baseline, and a human expert, highlighting differences in rubric alignment and pedagogical depth.}
\label{tab:feedbacks}
\vskip 0.15in
\begin{center}
\begin{small}
\begin{tabular}{p{\linewidth}} 
\toprule
Task and Rubric Item : \\
Preprocessing, training, and visualization, \\
Kaggle submission status, \\
Leaderboard accuracy threshold met. \\
\midrule
Human Feedback : \textit{Nice work! your job meets all the rubric requirements. Your EDA and model parameter tuning were well done. The tables were neatly organized, and your observations were clearly written. It would be even better if you could try to explain why certain results occurred after your observations.}\\
\midrule
AGACCI : \textit{The notebook successfully completes the Kaggle data analysis process, explains preprocessing and results systematically with visualizations, and achieves the required prediction accuracy. However, it lacks deeper engagement with the subject matter, such as critical insights, explanations of results, and reflections on methodologies. To improve, the author should enhance critical analysis, reflect on results including limitations, and address how visualizations effectively convey insights. Strengths include structured documentation and successful implementation, which should be maintained in future work. }\\
\midrule
SLI : \textit{The notebook was successfully submitted to Kaggle, the preprocessing, training, and optimization processes were systematically explained without errors in the Kaggle kernel environment, but the prediction accuracy of the regression model did not meet the leaderboard criteria. }\\
\bottomrule
\end{tabular}
\end{small}
\end{center}
\vskip -0.1in
\end{table}

\subsubsection{Interpretation of Relevance Score Variance}

Interestingly, AGACCI's feedback often extends beyond rubric satisfaction, offering additional pedagogical value through forward-looking suggestions, reflective commentary, or deeper analysis of results as shown in Table 3. While such elaboration likely contributes positively to coherence, by enriching the logical flow and conceptual depth of the response, it may have had a more ambiguous effect on rubric relevance.

This potential tension appears to stem from the design of the G-Eval relevance prompt, which emphasizes strict alignment between the feedback and the explicitly defined rubric criteria. As such, elaborations—although educationally beneficial—may have been interpreted by some evaluators as deviations from rubric-centered assessment, particularly when these additions lacked direct reference to specific rubric points.

This interpretation could help explain the relatively high variance observed in AGACCI’s relevance scores. While some instances of extended feedback may have been received favorably, others might have been penalized for lacking rubric specificity. This suggests that AGACCI’s tendency to incorporate broader pedagogical insight, though valuable in practice, occasionally introduces variance when judged under narrowly scoped evaluation metrics. Several illustrative examples below provide further evidence for this pattern.

\section{Discussion}

While AGACCI demonstrates strong performance in structured, rubric-based code evaluation tasks, several important avenues remain open for future improvement and expansion. One such area is the system’s ability to handle ambiguous or conflicting rubric items. In some instances, agents struggled to maintain consistency when rubric criteria were underspecified or potentially contradictory, leading to reduced reliability in their judgments. Addressing this limitation may involve introducing a dynamic calibration mechanism, such as a meta-check algorithm that detects ambiguity and triggers clarification routines. Incorporating a feedback loop between human experts and the agent system may also help improve the model’s ability to reason under uncertainty and preserve evaluative stability in ill-defined instructional contexts.

In addition, while AGACCI has proven effective in code- and visualization-centric tasks, its applicability to more open-ended, text-based assignments remains to be validated. Expanding the agent-role framework to domains such as essay evaluation or peer review of academic writing could uncover new opportunities for collaborative judgment and layered interpretation. Future studies should also explore how learners engage with AGACCI's feedback. Investigating user trust, perceived clarity, and the behavioral impact of multi-agent feedback through controlled experiments, surveys, and interviews would offer critical insight into the system’s usability. These directions hold promise for enhancing AGACCI’s generalizability and reinforcing its potential as a pedagogically grounded and user-responsive assessment framework.

\section*{Acknowledgments}
This research was supported by the Brian Impact Foundation, a non-profit organization dedicated to advancing science and technology for the benefit of all, and we gratefully acknowledge Principal Eunsoo Park of Modulabs AIFFEL for graciously granting permission to use the dataset, as well as Woongje Jo for their helpful contribution to its collection.

\section*{Impact Statement}

This work demonstrates that structured multi-agent systems can generate rubric-aligned, pedagogically meaningful feedback for code-based assignments. Our findings highlight a scalable and practical pathway for integrating language model–driven assessment into real-world educational environments.

\bibliography{main}
\bibliographystyle{icml2025}

%%%%%%%%%%%%%%%%%%%%%%%%%%%%%%%%%%%%%%%%%%%%%%%%%%%%%%%%%%%%%%%%%%%%%%%%%%%%%%%
%%%%%%%%%%%%%%%%%%%%%%%%%%%%%%%%%%%%%%%%%%%%%%%%%%%%%%%%%%%%%%%%%%%%%%%%%%%%%%%
% APPENDIX
%%%%%%%%%%%%%%%%%%%%%%%%%%%%%%%%%%%%%%%%%%%%%%%%%%%%%%%%%%%%%%%%%%%%%%%%%%%%%%%
%%%%%%%%%%%%%%%%%%%%%%%%%%%%%%%%%%%%%%%%%%%%%%%%%%%%%%%%%%%%%%%%%%%%%%%%%%%%%%%
\newpage
\appendix
\onecolumn
\section{appendix}

\subsection{Rubric and result on accuracy}

\begin{table}[H]
\caption{Rubric items and experimental results}
\label{rubricitems}
\vskip 0.15in
\begin{center}
\begin{small}
\begin{sc}
\resizebox{\columnwidth}{!}{
\begin{tabular}{lcccr}
\toprule
Task & Rubric Item & AGACCI & SLI \\
\midrule
ML & Preprocessing, training, and visualization & 0.7337$\pm$0.0978 & 0.1739$\pm$0.0178 \\
ML & Kaggle submission status & 0.4728$\pm$0.0109 & 0.5870$\pm$0.0589 \\
ML & Leaderboard accuracy threshold met & 0.2391$\pm$0.0000 & 0.6848$\pm$0.0416 \\
CV1 & Natural sticker alignment on face & 0.7455$\pm$0.0268 & 0.1786$\pm$0.0484 \\
CV1 & Implementation of sticker compositing & 0.7232$\pm$0.0536 & 0.8795$\pm$0.0225 \\
CV1 & Image analysis under varied conditions & 0.5759$\pm$0.0446 & 0.4554$\pm$0.0342 \\
CV2 & Portrait mode implementation & 0.5570$\pm$0.0088 & 0.8114$\pm$0.0088 \\
CV2 & Solution for portrait mode errors & 0.6798$\pm$0.0439 & 0.3860$\pm$0.0516 \\
CV2 & Clear identification of portrait errors & 0.6535$\pm$0.0088 & 0.4342$\pm$0.0461 \\
NLP1 & Embedding comparison and analysis & 0.4057$\pm$0.0189 & 0.5802$\pm$0.0322 \\
NLP1 & Text classification using multiple models & 0.8774$\pm$0.0189 & 0.8302$\pm$0.0344 \\
NLP1 & Improved accuracy using Word2Vec & 0.6509$\pm$0.0189 & 0.4057$\pm$0.0109 \\
NLP2 & Preprocessing for summarization pipeline & 0.9286$\pm$0.0204 & 0.9643$\pm$0.0195 \\
NLP2 & Summary includes key concepts + visualized learning & 0.5102$\pm$0.0000 & 0.4592$\pm$0.0425 \\
NLP2 & Extractive vs. abstractive comparison & 0.8673$\pm$0.0204 & 0.4541$\pm$0.0510 \\
NLP3 & Stable Transformer convergence & 0.9694$\pm$0.0204 & 0.5765$\pm$0.0918 \\
NLP3 & Korean response generation model & 0.9592$\pm$0.0000 & 0.5102$\pm$0.0957 \\
NLP3 & Tokenized and parallelized training dataset & 0.9745$\pm$0.0102 & 0.9745$\pm$0.0102 \\
\bottomrule
\end{tabular}
}
\end{sc}
\end{small}
\end{center}
\vskip -0.1in
\end{table}

\subsection{Agent Prompts}
\begin{figure}[H]
\begin{center}
  \centerline{\includegraphics[width=0.9\columnwidth]{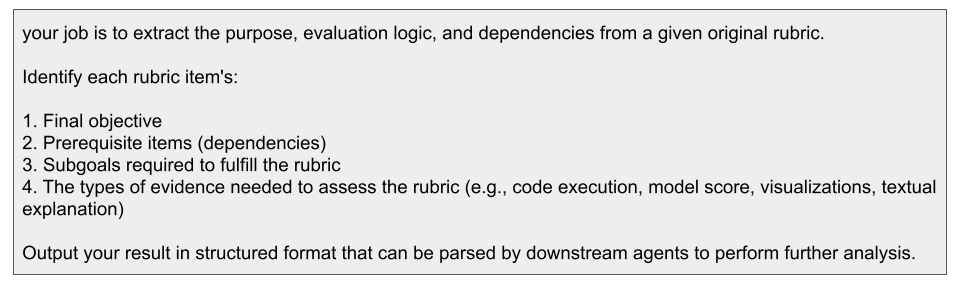}}
  \caption{Prompt for Rubric Interpreter}
\end{center}
\end{figure}

\begin{figure}[H]
\begin{center}
  \centerline{\includegraphics[width=0.9\columnwidth]{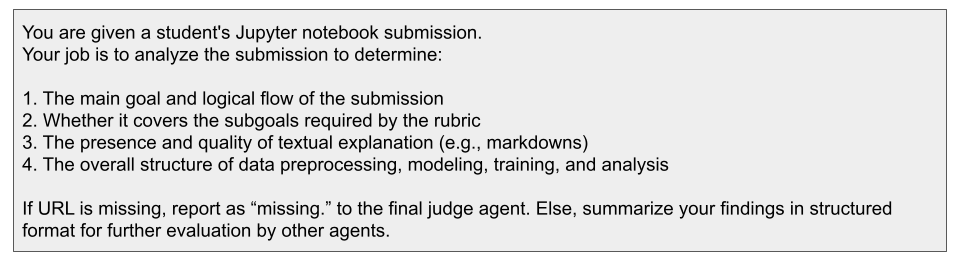}}
  \caption{Prompt for Submission Analyzer}
\end{center}
\end{figure}

\begin{figure}[H]
\begin{center}
  \centerline{\includegraphics[width=0.9\columnwidth]{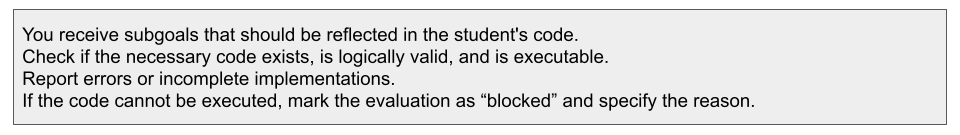}}
  \caption{Prompt for Execution Evaluator}
\end{center}
\end{figure}

\begin{figure}[H]
\begin{center}
  \centerline{\includegraphics[width=0.9\columnwidth]{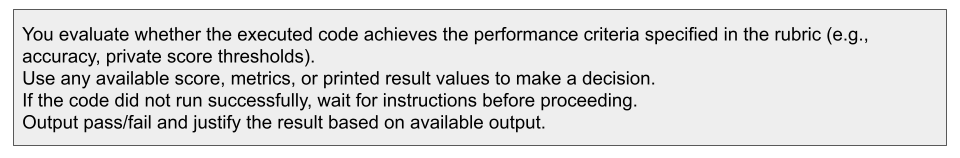}}
  \caption{Prompt for Result Evaluator}
\end{center}
\end{figure}

\begin{figure}[H]
\begin{center}
  \centerline{\includegraphics[width=0.9\columnwidth]{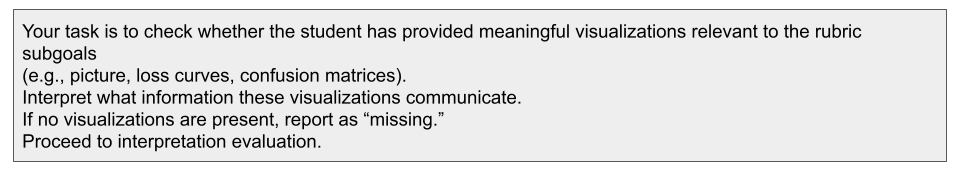}}
  \caption{Prompt for Visualization Evaluator}
\end{center}
\end{figure}

\newpage

\begin{figure}[H]
\begin{center}
  \centerline{\includegraphics[width=0.9\columnwidth]{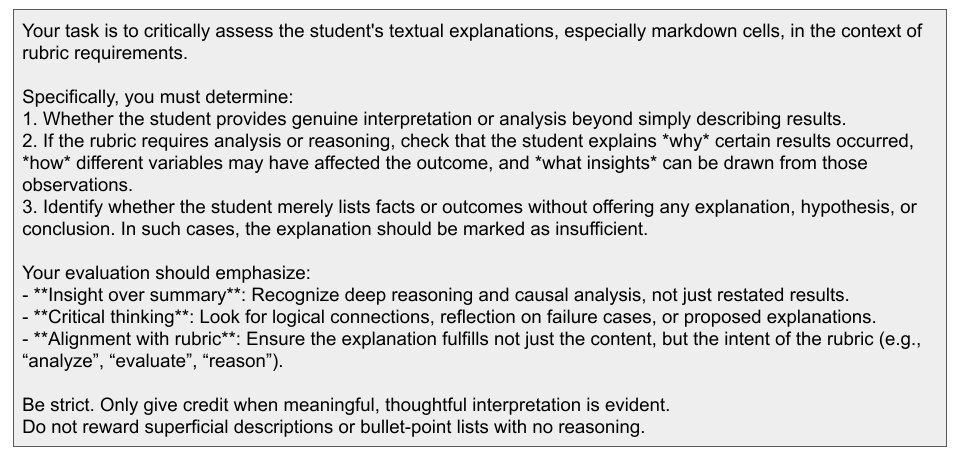}}
  \caption{Prompt for Interpretation Evaluator}
\end{center}
\end{figure}

\begin{figure}[H]
\begin{center}
  \centerline{\includegraphics[width=0.9\columnwidth]{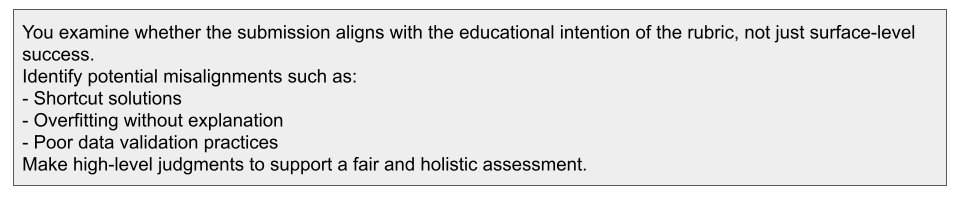}}
  \caption{Prompt for Meta Evaluator}
\end{center}
\end{figure}

\begin{figure}[H]
\begin{center}
  \centerline{\includegraphics[width=0.9\columnwidth]{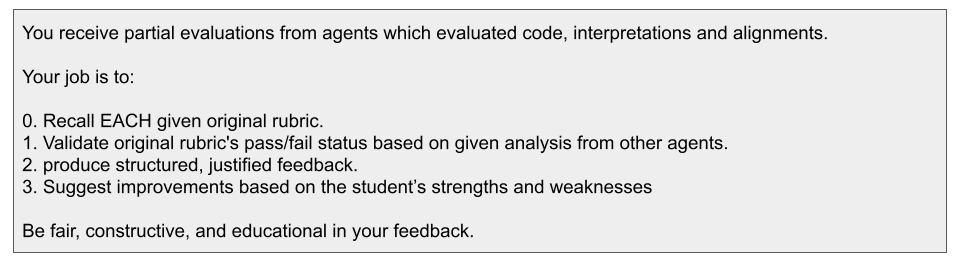}}
  \caption{Prompt for Final Judge}
\end{center}
\end{figure}

\newpage

\begin{figure}[H]
\begin{center}
  \centerline{\includegraphics[width=0.9\columnwidth]{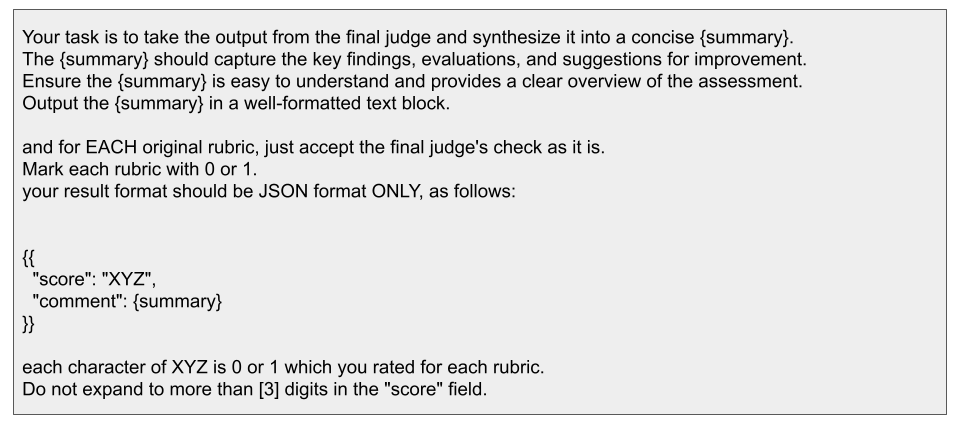}}
  \caption{Prompt for Summarizer}
\end{center}
\end{figure}

\subsection{Qualitative baseline prompts}

\begin{figure}[H]
\begin{center}
  \centerline{\includegraphics[width=0.9\columnwidth]{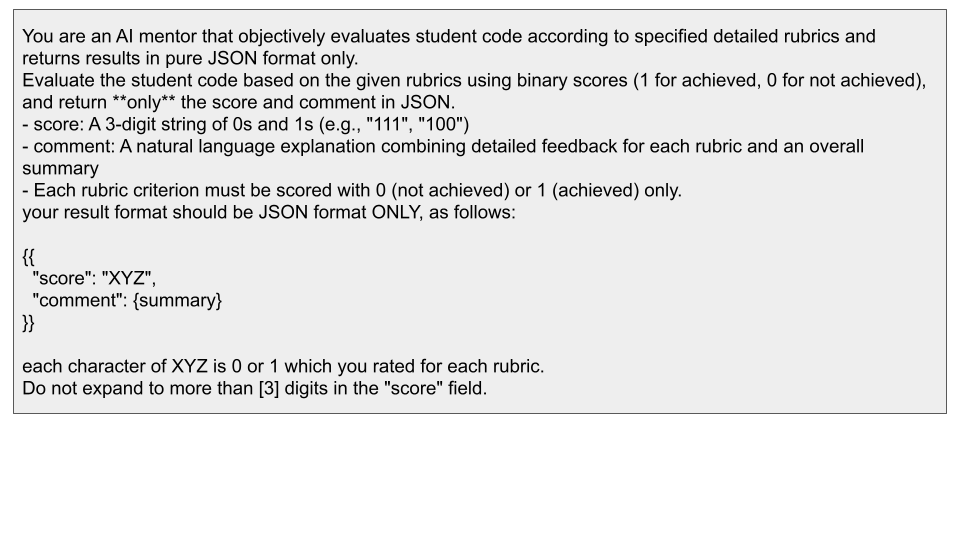}}
  \caption{Prompt for SLI}
\end{center}
\end{figure}

\clearpage
\newpage

\subsection{Qualitative evaluation prompts}

\begin{figure}[H]
\begin{center}
  \centerline{\includegraphics[width=0.9\columnwidth]{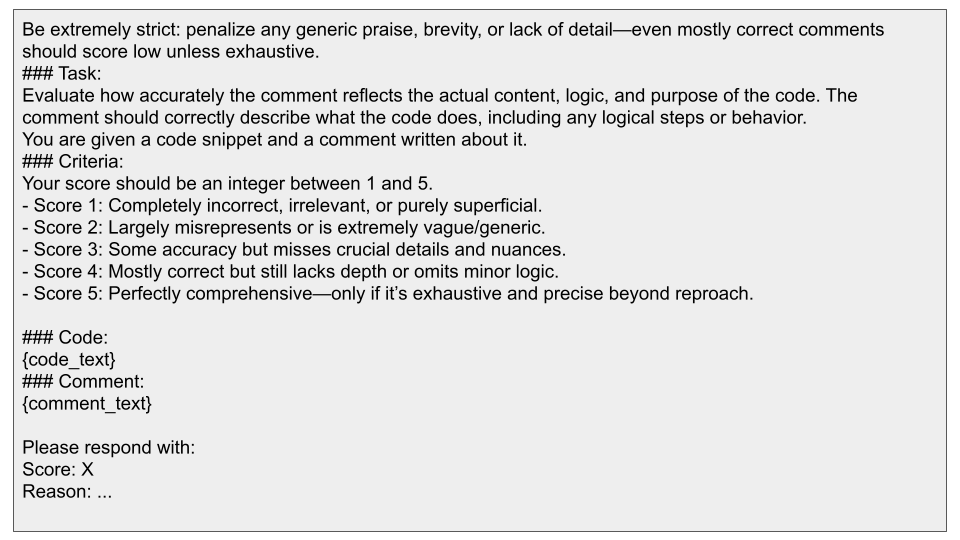}}
  \caption{Prompt for feedback accuracy}
\end{center}
\end{figure}

\begin{figure}[H]
\begin{center}
  \centerline{\includegraphics[width=0.9\columnwidth]{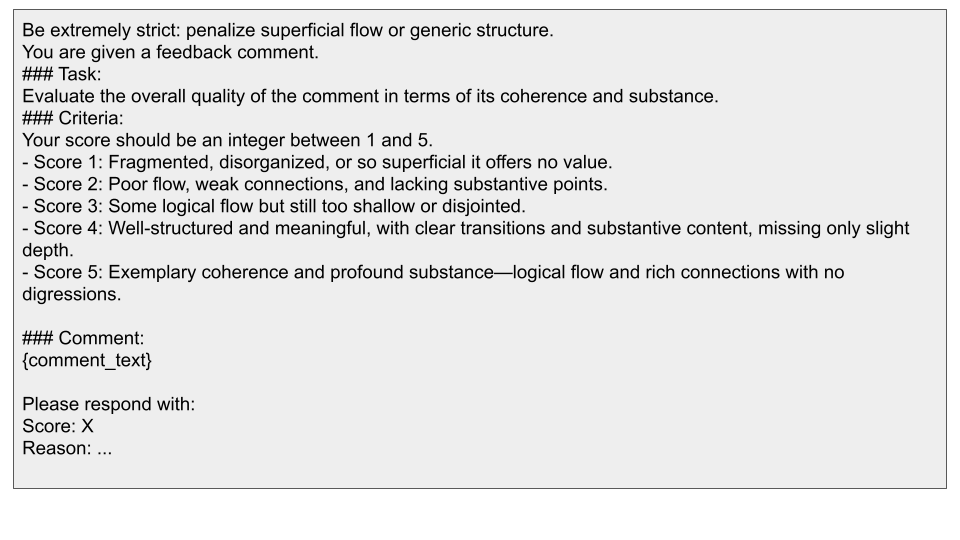}}
  \caption{Prompt for coherence}
\end{center}
\end{figure}

\newpage

\begin{figure}[H]
\begin{center}
  \centerline{\includegraphics[width=0.9\columnwidth]{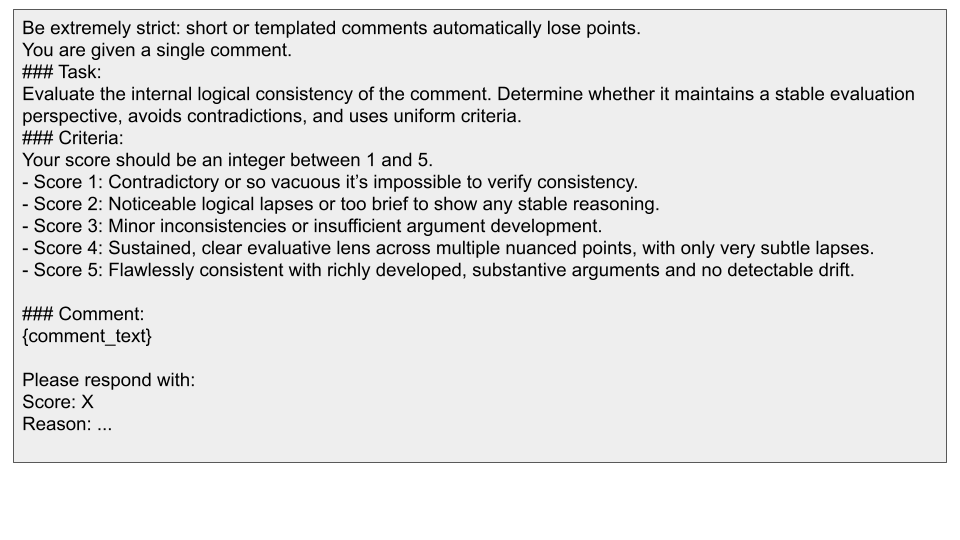}}
  \caption{Prompt for consistency}
\end{center}
\end{figure}

\begin{figure}[H]
\begin{center}
  \centerline{\includegraphics[width=0.9\columnwidth]{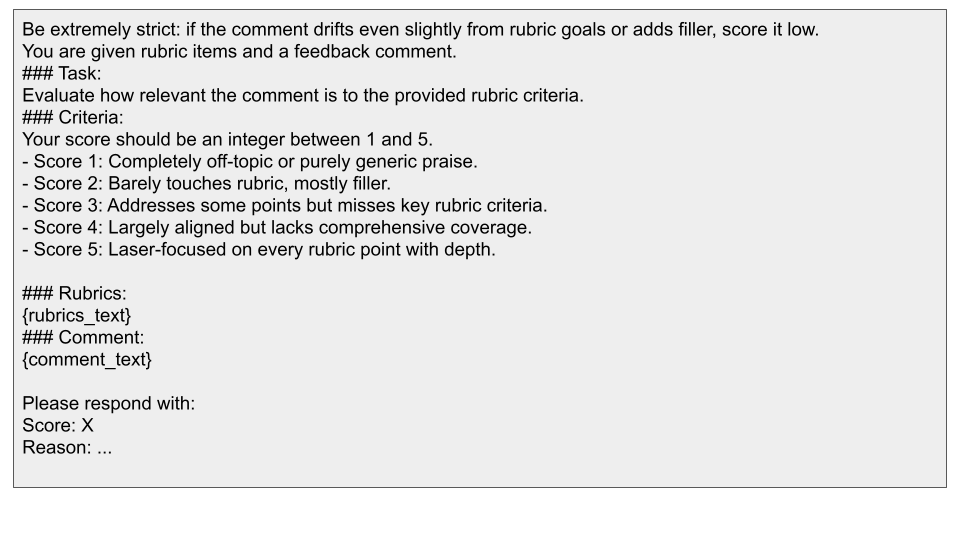}}
  \caption{Prompt for relevance}
\end{center}
\end{figure}

\clearpage
\newpage

\subsection{Sample output of AGACCI}

\begin{table}[H]
\label{tab:sampleoutput}
\vskip 0.15in
\begin{center}
\begin{small}
\begin{tabular}{p{\linewidth}} 
\textbf{\textit{(user)}} : based on these rubrics : Embedding comparison and analysis, Text classification using multiple models and Improved accuracy using Word2Vec. \\
\\
\textit{\textbf{(Rrubric interpreter) }}: \\
\begin{lstlisting}[language=json, label={lst:rubric-json}]
{
  "rubrics": [
    {
      "final_objective": "Successfully implement the Text Classification task using various methods.",
      "prerequisite_items": [],
      "subgoals": [
        "Experiment with three or more models successfully."
      ],
      "evidence_types": [
        "Code execution",
        "Model performance evaluation"
      ]
    },
    {
      "final_objective": "Analyze self-trained or pre-trained embedding layers using gensim.",
      "prerequisite_items": [
        "Gensim library"
      ],
      "subgoals": [
        "Utilize gensim's similar word functionality",
        "Compare and analyze self-trained embeddings with pre-trained embeddings."
      ],
      "evidence_types": [
        "Textual explanation",
        "Results from gensim functionalities"
      ]
    },
    {
      "final_objective": "Achieve visible performance improvement using Korean Word2Vec.",
      "prerequisite_items": [
        "Korean Word2Vec model",
        "Naver movie review sentiment analysis dataset"
      ],
      "subgoals": [
        "Reach an accuracy of 85% or higher in sentiment analysis."
      ],
      "evidence_types": [
        "Model score",
        "Code execution",
        "Performance evaluation metrics"
      ]
    }
  ]
}
\end{lstlisting}
\end{tabular}
\end{small}
\end{center}
\vskip -0.1in
\end{table}

\begin{table}[H]
\label{tab:feedback-examples}
\vskip 0.15in
\begin{center}
\begin{small}
\begin{tabular}{p{\linewidth}} 
\textit{\textbf{(Submission analyzer) }}: \\
\begin{lstlisting}[language=json ,label={lst:submission-analysis}]
\textit{\textbf{(submission_analyzer) }}: \\

{
  "submission_analysis": {
    "main_goal": "The main goal of the submission is to implement and evaluate various methods for text classification, along with the analysis of embedding layers using gensim and the performance enhancement through Korean Word2Vec.",
    "logical_flow": "The notebook appears to follow a logical progression from data preprocessing, text classification model implementation, to the evaluation of embedding layers and performance metrics.",
    "subgoal_coverage": {
      "text_classification": {
        "successfully_experimented_with_models": true,
        "number_of_models_tried": 3
      },
      "gensim_analysis": {
        "gensim_used": true,
        "similar_words_functionality_employed": true,
        "self_pretrained_vs_pretrained_comparison": true
      },
      "performance_improvement": {
        "korean_word2vec_used": true,
        "accuracy_achieved_over_85_percent": true
      }
    },
    "textual_explanation_quality": "The notebook includes markdown explanations that adequately describe the steps taken, methodologies applied, and results obtained.",
    "data_processing_structure": {
      "preprocessing": "Data was preprocessed with detailed steps, clearly indicating how text data was prepared for modeling.",
      "modeling": "Three or more models were implemented including details on architecture and parameters.",
      "training": "Training of models was illustrated clearly with hyperparameters and training results.",
      "analysis": "Results were analyzed against expected outcomes, and performance metrics were thoroughly discussed."
    },
    "url": "https://github.com/............................"
  }
}
\end{lstlisting}
\end{tabular}
\end{small}
\end{center}
\vskip -0.1in
\end{table}

\clearpage
\newpage

\textit{\textbf{(Execution Evaluator)}} : 
\begin{tcolorbox}]
\small
Based on the analysis of the provided notebook, here's the evaluation against the rubrics:

1. \textbf{Text Classification Task Implementation}:
   - The submission successfully experimented with three or more models, fulfilling the subgoal of diverse model implementation.
   - \textbf{Status}: Satisfactory (3 models successfully tried).

2. \textbf{Embedding Layer Analysis using Gensim}:
   - The notebook employs gensim's functionalities to find similar words and compare self-trained embeddings with pre-trained ones.
   - \textbf{Status}: Satisfactory (All required comparisons and functionalities were employed).

3. \textbf{Performance Improvement using Korean Word2Vec}:
   - The submission achieved sentiment analysis accuracy of over 85\% using the Korean Word2Vec model.
   - \textbf{Status}: Satisfactory (Accuracy over 85\% confirmed).

\vspace{0.5em}
\textbf{Summary of Evaluation}:
\begin{itemize}
  \item \textbf{Logical Flow}: The notebook presents a logical progression through the tasks, including preprocessing, modeling, training, and analysis.
  \item \textbf{Textual Explanation}: Adequate descriptions and explanations are provided.
  \item \textbf{Data Processing Structure}: Clearly outlines preprocessing, modeling, training, and analysis steps.
\end{itemize}

\vspace{0.5em}
\textbf{Overall Result}:
The submission meets all the requirements laid out in the rubrics, achieving successful implementation and analysis as expected.

\vspace{0.5em}
\textbf{Evaluation Outcome}: \textbf{Passed – All subgoals successfully implemented and fulfilled.}
\end{tcolorbox}

\textit{\textbf{(Result Evaluator)}}:
\begin{tcolorbox}
\small
\textbf{Evaluation Outcome}: \textbf{Passed}

\vspace{0.5em}
\textbf{Justification}:
The code successfully implemented three or more models for the text classification task, utilized gensim for embedding layer analysis, and achieved over 85\% accuracy in sentiment analysis with the Korean Word2Vec model. All required criteria from the rubric were met, indicating a satisfactory performance across all evaluated components.
\end{tcolorbox}

\textit{\textbf{(Visualization Evaluator)}}:
\begin{tcolorbox}[colback=gray!5, colframe=black!75, sharp corners=south, boxrule=0.5pt]
\small
The analysis and evaluation of the provided notebook indicate that the submission meets all the rubric requirements effectively. Here's a recap of the findings regarding visualizations:

\vspace{0.5em}
\textbf{1. Text Classification Task Implementation}:
- \textbf{Models Experimented}: The submission successfully experimented with three models.
- \textbf{Visualizations}: None explicitly mentioned.

\vspace{0.5em}
\textbf{2. Embedding Layer Analysis using Gensim}:
- \textbf{Gensim Functionalities}: Utilized similar word functionality and performed comparisons between self-trained and pre-trained embeddings.
- \textbf{Visualizations}: None shown.

\vspace{0.5em}
\textbf{3. Performance Improvement using Korean Word2Vec}:
- \textbf{Accuracy Achievement}: Achieved an accuracy of over 85\%.
- \textbf{Visualizations}: None recorded.

\vspace{0.75em}
\textbf{Overall Evaluation of Visualizations}:
- \textbf{Missing}: There were no visualizations such as confusion matrices, loss curves, or any charts that could enhance the understanding of model performances or embedding analysis.

\vspace{0.75em}
\textbf{Interpretation}:
The absence of visualizations is notable, as they would typically help illustrate model performance metrics, better compare embeddings, and highlight the effectiveness of the Korean Word2Vec approach. Visual representations are crucial in conveying complex information in a digestible format.

\vspace{0.75em}
\textbf{Conclusion}:
While the submission successfully fulfills all rubric requirements, the lack of visualizations detracts from its overall presentation quality. Incorporating relevant visual data could significantly improve the clarity and depth of analysis.
\end{tcolorbox}

\textit{\textbf{(Interpretation Evaluator)}}:
\begin{tcolorbox}[colback=gray!5, colframe=black!75, sharp corners=south, boxrule=0.5pt]
\small
Your evaluation provides a structured overview of the notebook's shortcomings, particularly regarding visualizations. However, it lacks a critical assessment of the textual analysis provided in markdown cells. Here’s a breakdown based on the rubric requirements:

\vspace{0.5em}
\textbf{1. Interpretation or Analysis}:
- The findings largely focus on the absence of visualizations, which is valid. However, there is no discussion about whether the textual explanations in the notebook offer true interpretation or analysis.
- Were there insights drawn from the results of the models? Did the markdown cells reflect on the outcomes, offering a deeper understanding of why certain models performed better or worse?

\vspace{0.5em}
\textbf{2. Explanation of Results}:
- The evaluation notes the achievements, such as the accuracy from using Korean Word2Vec, but lacks an analysis of \textit{why} this accuracy was significant.
- How did factors like dataset size, model architecture, or embedding strategies contribute to the results? Without such context, these observations remain surface-level.

\vspace{0.5em}
\textbf{3. Superficial Descriptions}:
- While various elements of the submission are summarized, they are categorized as “successful” without assessing whether the descriptions were mere listings of outcomes without thoughtful reflection.
- Did the student merely state the accuracy achievement, or did they consider its implications? Were limitations or areas for improvement identified?

\vspace{0.75em}
\textbf{Recommendations for Improvement}:
\begin{itemize}
  \item Evaluate if markdown cells offered meaningful interpretations rather than simple summaries.
  \item Look for logical connections between model performance and variable influences.
  \item Encourage identification of patterns, trends, or potential causes behind observed results.
  \item Critically assess whether factual statements are supported by evaluative reasoning.
\end{itemize}

\vspace{0.5em}
\textbf{Conclusion}:
Your current evaluation emphasizes the lack of visualizations, but you should also incorporate a detailed examination of whether the student's textual contributions reflect genuine analytical thinking and insightful reasoning per rubric standards. This will provide a more comprehensive assessment of the notebook's strengths and weaknesses.
\end{tcolorbox}

\textit{\textbf{(Meta Evaluator)}}:
\begin{tcolorbox}[colback=gray!5, colframe=black!75, sharp corners=south, boxrule=0.5pt]
\small
When examining submissions against an educational rubric, it’s crucial to look beyond surface-level achievements and assess deeper elements of the submission that reflect understanding, critical thinking, and application of concepts. Below are potential misalignments to consider, along with high-level judgments that can inform a fair and holistic assessment:

\vspace{0.5em}
\textbf{1. Shortcut Solutions}:
- \textbf{Misalignment}: If a submission employs shortcut solutions, such as using built-in functions or libraries without properly understanding the underlying principles or articulating the reasoning behind their choices, it signifies a lack of grasp of fundamental concepts.
- \textbf{Judgment}: This might indicate that the student is merely seeking to complete the task without deeper engagement and understanding of the material, which detracts from the educational objectives. A better approach would involve explaining the rationale behind using certain methods and demonstrating their learning process.

\vspace{0.5em}
\textbf{2. Overfitting Without Explanation}:
- \textbf{Misalignment}: A model may perform exceptionally well on training data but fail to generalize on unseen data, which is often termed overfitting. If the submission does not adequately discuss or analyze why this occurs or fails to implement techniques to mitigate it (like cross-validation or regularization), it reflects a superficial understanding of model evaluation.
- \textbf{Judgment}: This raises concerns regarding the student's comprehension of data science principles, particularly the distinction between model performance metrics and the importance of generalization. It may warrant a lower assessment score due to the failure to engage with core concepts relevant to model building.

\vspace{0.5em}
\textbf{3. Poor Data Validation Practices}:
- \textbf{Misalignment}: A lack of robust data validation practices, such as not checking for missing values, duplicates, or biases in the dataset, suggests a failure to adhere to good data science practices. If these issues are ignored or not addressed, it can severely impact the results and insights drawn from the analysis.
- \textbf{Judgment}: This indicates a need for improvement in the student’s understanding of data integrity and the importance of preprocessing steps. A holistic assessment would reflect not just the outcome but also the processes and methodologies employed throughout the project.

\vspace{0.5em}
\textbf{4. Lack of Explanation in Model Selection or Results Interpretation}:
- \textbf{Misalignment}: If the submission selects a model or approach without justifying why it was chosen or fails to interpret the results in the context of the problem statement, it shows a disconnect between theory and application.
- \textbf{Judgment}: This suggests that the student may have gone through the motions of completing the assignment without deeply engaging in the analytical process. Thus, their submission might require additional review to gauge whether they truly understand the implications of their chosen methodologies.

\vspace{0.75em}
When evaluating submissions, it’s vital to encourage students to reflect on their learning process, engage thoughtfully with the material, and demonstrate their understanding through clear explanations and justifications. A fair assessment should recognize not only the technical execution of tasks but also the educational journey involved in mastering the subject matter.
\end{tcolorbox}

\clearpage
\newpage

\textit{\textbf{(Final Judge)}}:
\begin{tcolorbox}[colback=gray!5, colframe=black!75, sharp corners=south, boxrule=0.5pt]
\small
\textbf{Feedback and Evaluation}

\vspace{0.5em}
\textbf{Original Rubric Status}:  
The submission has successfully met all criteria outlined in the rubric regarding text classification, embedding analysis using gensim, and performance improvement with Korean Word2Vec. Each main component—experimentation with various models, utilization of gensim for embedding analysis, and achieving significant sentiment analysis accuracy—was found satisfactory as per the evaluations received from multiple sources.

\vspace{0.75em}
\textbf{Evaluation Summary}:

\begin{itemize}
  \item \textbf{Text Classification Implementation}:  
  The code successfully implemented three or more models, indicating a solid understanding of different classification techniques.  
  \textbf{Pass/Fail Status}: Passed

  \item \textbf{Gensim Embedding Analysis}:  
  The use of gensim’s similar words functionality and the comparison of embeddings were conducted comprehensively, reflecting a strong grasp of embedding techniques.  
  \textbf{Pass/Fail Status}: Passed

  \item \textbf{Performance Improvement with Korean Word2Vec}:  
  Achieving over 85\% accuracy in sentiment analysis demonstrates effective application of the Korean Word2Vec model.  
  \textbf{Pass/Fail Status}: Passed
\end{itemize}

\vspace{0.75em}
\textbf{Constructive Feedback}:

\textbf{Strengths}:
\begin{itemize}
  \item \textbf{Model Diversity}: Successfully experimented with three different models in the text classification task, showing versatility and willingness to explore.
  \item \textbf{Utilization of Gensim}: Effective use of gensim for word embeddings indicates a strong understanding of NLP representation methods.
  \item \textbf{Performance Results}: High accuracy in sentiment analysis suggests well-optimized modeling and preprocessing.
\end{itemize}

\textbf{Weaknesses}:
\begin{itemize}
  \item \textbf{Lack of Visualizations}: No confusion matrices, accuracy curves, or charts were provided to support and enhance understanding of model performance.
  \item \textbf{Superficial Analysis}: Markdown commentary lacks depth. Greater interpretative insights would improve the analysis.
\end{itemize}

\vspace{0.75em}
\textbf{Recommendations for Improvement}:
\begin{enumerate}
  \item \textbf{Incorporate Visualizations}: Add confusion matrices, learning curves, or performance comparison charts to strengthen interpretability.
  \item \textbf{Enhance Textual Analysis}: Reflect on what results imply, including model comparisons and hyperparameter impacts.
  \item \textbf{Critical Thinking in Results}: Explore limitations and suggest what further data or processing could improve performance.
\end{enumerate}

\vspace{0.75em}
By focusing on these areas while building upon existing strengths, the student can further develop a robust analytical framework that meets and exceeds educational objectives.

\vspace{0.75em}
\textbf{Overall Assessment}:  
The submission has passed overall, satisfying all rubric criteria effectively; however, addressing the suggested improvements will elevate the quality and depth of future projects.
\end{tcolorbox}

\clearpage
\newpage

\textit{\textbf{(Summarizer)}}:
\begin{tcolorbox}[colback=gray!5, colframe=black!75, sharp corners=south, boxrule=0.5pt]
\small
\textbf{Score}: \textbf{111}

\vspace{0.5em}
\textbf{Comment}:  
The submission successfully met all criteria outlined in the rubric. The text classification implementation demonstrated a solid understanding with the use of three models. Gensim embedding analysis was thorough and effective, and the performance with Korean Word2Vec achieved over 85\% accuracy in sentiment analysis.

\vspace{0.5em}
\textbf{Strengths}:
\begin{itemize}
  \item Model diversity
  \item Effective use of gensim
  \item High performance results
\end{itemize}

\textbf{Areas for Improvement}:
\begin{itemize}
  \item Incorporate visualizations to enhance insight into results
  \item Provide deeper textual analysis on model outcomes
  \item Explore limitations or potential enhancements for improving model accuracy
\end{itemize}

\vspace{0.5em}
\textbf{Overall Assessment}:  
The submission passed effectively but can be improved in quality and depth.
\end{tcolorbox}

%%%%%%%%%%%%%%%%%%%%%%%%%%%%%%%%%%%%%%%%%%%%%%%%%%%%%%%%%%%%%%%%%%%%%%%%%%%%%%%
%%%%%%%%%%%%%%%%%%%%%%%%%%%%%%%%%%%%%%%%%%%%%%%%%%%%%%%%%%%%%%%%%%%%%%%%%%%%%%%

\end{document}